\documentclass[authoryear,a4paper,10pt,onecolumn,preprint]{elsarticle}

\usepackage{mathptmx}
\usepackage{amsmath,amsfonts,amssymb}
\usepackage{makeidx}
\usepackage{graphicx}
\usepackage{epstopdf}
\usepackage{multirow}
\usepackage{rotating}
\usepackage{color}
\usepackage{tablefootnote}
\usepackage{lscape}
\usepackage{subfigure}
\usepackage{tablefootnote}
\usepackage[T1]{fontenc}
\usepackage{aecompl}
\usepackage[margin=1.5cm]{geometry}

\usepackage{listings}
 
\definecolor{codegreen}{rgb}{0,0.6,0}
\definecolor{codegray}{rgb}{0.5,0.5,0.5}
\definecolor{codepurple}{rgb}{0.58,0,0.82}
\definecolor{backcolour}{rgb}{0.95,0.95,0.92}
 
\lstdefinestyle{mystyle}{
    backgroundcolor=\color{backcolour},   
    commentstyle=\color{codegreen},
    keywordstyle=\color{magenta},
    numberstyle=\tiny\color{codegray},
    stringstyle=\color{codepurple},
    basicstyle=\footnotesize,
    breakatwhitespace=false,         
    breaklines=true,                 
    captionpos=b,                    
    keepspaces=true,                 
    numbers=left,                    
    numbersep=5pt,                  
    showspaces=false,                
    showstringspaces=false,
    showtabs=false,                  
    tabsize=2
}

\journal{Astronomy and Computing}

\newcommand{\comest}{\texttt{ComEst}}
\newcommand{\se}{\texttt{SExtractor}}
\newcommand{\fits}{\textrm{FITS}}
\newcommand{\galsim}{\texttt{GalSim}}
\newcommand{\zp}{\textrm{ZP}}
\newcommand{\bcs}{\textrm{BCS}}
\newcommand{\python}{\texttt{Python}}

\newcommand{\fcom}{\ensuremath{\mathrm{f_{\mathrm{com}}}}}
\newcommand{\fpur}{\ensuremath{\mathrm{f_{\mathrm{pur}}}}}
\newcommand{\fdr}{\ensuremath{\mathrm{f_{\mathrm{dr}}}}}

\newcommand{\percent}{\ensuremath{\%}}
\begin{document}

% use two columns
%\twocolumn[{ 
% front matter setting
\begin{frontmatter}

%\title{\comest: a Completeness Estimator of Charge-Coupled Device Images}
\title{\comest: a Completeness Estimator of Source Extraction on Astronomical Imaging}

\author[add1,add2]{I-Non Chiu}
\ead{inonchiu@usm.lmu.de}
\author[add2]{Shantanu Desai}
\ead{shantanu@usm.lmu.de}
\author[add3]{Jiayi Liu}
\ead{jiayiliu@usm.lmu.de}
\address[add1]{
Faculty of Physics, Ludwig-Maximilians University, Scheinerstr. 1, D-81679 M\"{u}nchen, Germany 
}
\address[add2]{
Excellence Cluster Universe, Boltzmannstr.\ 2, 85748 Garching, Germany 
}
\address[add3]{
Bosch Research and Technology Center North America,
4005 Miranda Ave \#200, Palo Alto, CA 94304, United States
}

\begin{abstract}

The completeness of  source detection is critical for analyzing the photometric and spatial properties of the population of interest observed by astronomical imaging.
We present a software package \comest, which calculates the completeness of source detection on charge-coupled device (CCD) images of astronomical observations, especially for the optical and near-infrared (NIR) imaging of galaxies and point sources.
The completeness estimator \comest\ is designed for the source finder \se\ used on the CCD images saved in the Flexible Image Transport System (\fits) format.
Specifically, \comest\ estimates the completeness of the source detection by deriving the detection rate of synthetic point sources and galaxies simulated on the observed CCD images.
In order to capture any observational artifacts or noise properties while deriving the completeness, \comest\ directly carries out the detection of simulated sources on the observed images.
Given an observed CCD image saved in  \fits\ format, \comest\ derives the completeness of the source detection from end to end as a function of source flux (or magnitude) and CCD position.
In addition, \comest\ can also estimate the purity of the source detection by comparing the catalog of the detected sources to the input catalogs of the simulated sources.
We run \comest\ on the images from Blanco Cosmology Survey (\bcs) and compare the derived completeness as a function of magnitude to the limiting magnitudes derived by using the Signal-to-Noise ratio (SNR) and number count histogram of the detected sources.
\comest\ is released as a \python\ package with an easy-to-use syntax and is publicly available at https://github.com/inonchiu/ComEst .

\end{abstract}


\end{frontmatter}

%}]

%%%%%%%%%%%%%%%%%%%%%%%%%%%%
%
% Introduction
%
%%%%%%%%%%%%%%%%%%%%%%%%%%%%
\section{Introduction}
\label{sec:introduction}

The optical and Near-Infrared (NIR) imaging provides one of the backbones of astronomy.
Since the early twentieth century, countless milestones have been made in various fields of astronomy based upon the data taken by the optical sky surveys, e.g., SDSS \citep[][]{york2000}.
Nowadays wide field optical/NIR surveys-- such as the Pan-Starrs \citep{morgan14}, DES \citep{flaugher05}, KiDS \citep{dejong12}, HSC \citep{miyazaki12}, 2MASS \citep{skrutskie06}, WISE \citep{wright10} and ATLAS \citep{shanks15}-- and dedicated deep imaging  (e.g., COSMOS \citep{koekemoer07} or HUDF \citep{beckwith06}) have become the frontier of astronomical studies in various topics.
In the next decade, the upcoming surveys-- for instance, LSST \citep{ivezic08} or \textit{Euclid} \citep{laureijs11}-- with the unprecedented deep imaging of large portion of the sky will revolutionize our understanding of the Universe.

Nowadays, modern astronomical observations are imaged by charge-coupled devices (CCD) saved in the Flexible Image Transport System (\fits) format, which various source finders, e.g., \citep[][]{bertin1996} are run on for detecting sources.
To analyze the observed images taken by various telescopes, the astronomical objects are identified by the source finder, and then  information about the photometric and spatial properties is extracted from the image.
It is therefore critical to verify the source detection on the observed image for the analysis of, for instance, modeling the luminosity function or spatial clustering of galaxies.
One of the most important factors for such studies is to quantify the completeness of the source detection on the observed images.

The completeness of the source detection indicates the fraction of objects present in the image, which can be detected by the source finder above a certain detection threshold against the noise.
The completeness is expected to decrease with increasing detection threshold because the required signal-to-noise ratio (SNR) of the source is higher.
At the same time, the purity of the source detection-- the fraction of the detected sources which are not spuriously detected by the source finder-- becomes higher due to the high SNR as well.
The lack of understanding the completeness (or purity) of the source detection could lead to  biased scientific results.
For example, the modeling of luminosity function at the faint end would be biased if one does not pay special care to the completeness near the detection threshold.

There are several ways to quantify the completeness of the source detection in terms of limiting depth (or limiting magnitude).
In general, one can compare the source catalog, which is extracted from the observed image by the source finder, to a reference catalog built from the image with deeper depth.
By comparing the source counts normalized by the observed area between two catalogs, one can calculate the completeness as the ratio of source counts given a limiting magnitude.
For example, the source count-magnitude relation assuming a power law, with the power law index estimated from the reference catalog can be fitted to the source catalog to determine the completeness, which is defined as the ratio of the number counts of the source catalog to the best-fit power law model at given magnitude \citep[e.g., ][]{zenteno11, chiu16b}.
However, this approach ignores the cosmic variance of the source properties on the sky and is frequently affected by the systematics between the catalogs, such as the difference in the observed filter systems, detection algorithms or the observational conditions. 
Another example is to link the photometric uncertainties of the detected sources to the completeness as a function of the survey depth \citep{rykoff15}.
Nevertheless, this method requires  intensive modeling of the sky noise and photometry measurement; in addition, the large number of  extracted sources in the overlapping region between the source and reference catalogs is needed in order to obtain the precise completeness estimates.

On the other hand, the most direct way to estimate the completeness of the source detection is to quantify the performance of the source finder by running the same detection algorithm on the objects simulated on the observed image.
In this way, the completeness (and purity) of the source detection can be derived by comparing the extracted catalog to the catalog of simulated sources used as the input of the simulation.
Simulating the sources on the observed image preserves any observational artifacts and sky noise while quantifying the performance of the source detection. 
Moreover, carrying out exactly the same algorithm to detect the simulated sources on the observed image provides an end-to-end verification and prevents any systematics arising from the catalog comparison. 
Several packages exist which can estimate the completeness using a similar methodology mentioned above, for example DAOPHOT \citep[][]{stetson1987} or 2DPHOT \citep[][]{labarbera08}.
However, DAOPHOT is specifically designed for  stellar-like objects and is not optimized for the extended sources such as galaxies.
For the latter example, 2DPHOT requires  intensive pre-modeling of the source properties (e.g., the morphology).
In addition, the packages mentioned above are less user-friendly compared to the other image simulation software, such as \galsim\ \citep{rowe15}. 
A more straightforward way for source simulation is to simulate the sources based on the models with various properties, so  that we have  full control and are independent from the observed sources.
%SD: got rid of the before full control

In this paper, we present the user-friendly software package \comest-- which is designed for estimating the completeness of the source detection on the observed image saved in the FITS format, by deriving the detection rate of the sources simulated with various properties assigned by the users.
In addition, \comest\ is designed for the source finder \se\ \citep{bertin1996}, which is widely used in the astronomy community for detecting the sources observed by the optical/NIR imaging.
This paper is organized as follows.
The methodology of \comest\ is described in Section~\ref{sec:method}, while we demonstrate the usage of \comest\ with an example and present the results in Section~\ref{sec:example_results}.
We conclude  made in Section~\ref{sec:conclusion}.

%%%%%%%%%%%%%%%%%%%%%%%%%%%%
%
%  Methods
%
%%%%%%%%%%%%%%%%%%%%%%%%%%%%

\section{Methods}
\label{sec:method}

\comest\ is developed as a completeness estimator for the  \se\ cataloging program  for any given imaging of point sources or galaxies with  well-calibrated photometric zeropoint (\zp).
To capture any observational effects which are already present in the image, \comest\ directly simulates sources with various properties (e.g., the sizes or fluxes) and runs \se\ to carry out the source detection on the observed image.
Therefore, by design, \comest\ heavily relies on the source finder \se\ and the image simulation toolkit \galsim\ \citep{rowe15}, which is used as the engine for simulating sources.

\begin{figure*}
\centering
\resizebox{\columnwidth}{!}{
\includegraphics[scale=1.0]{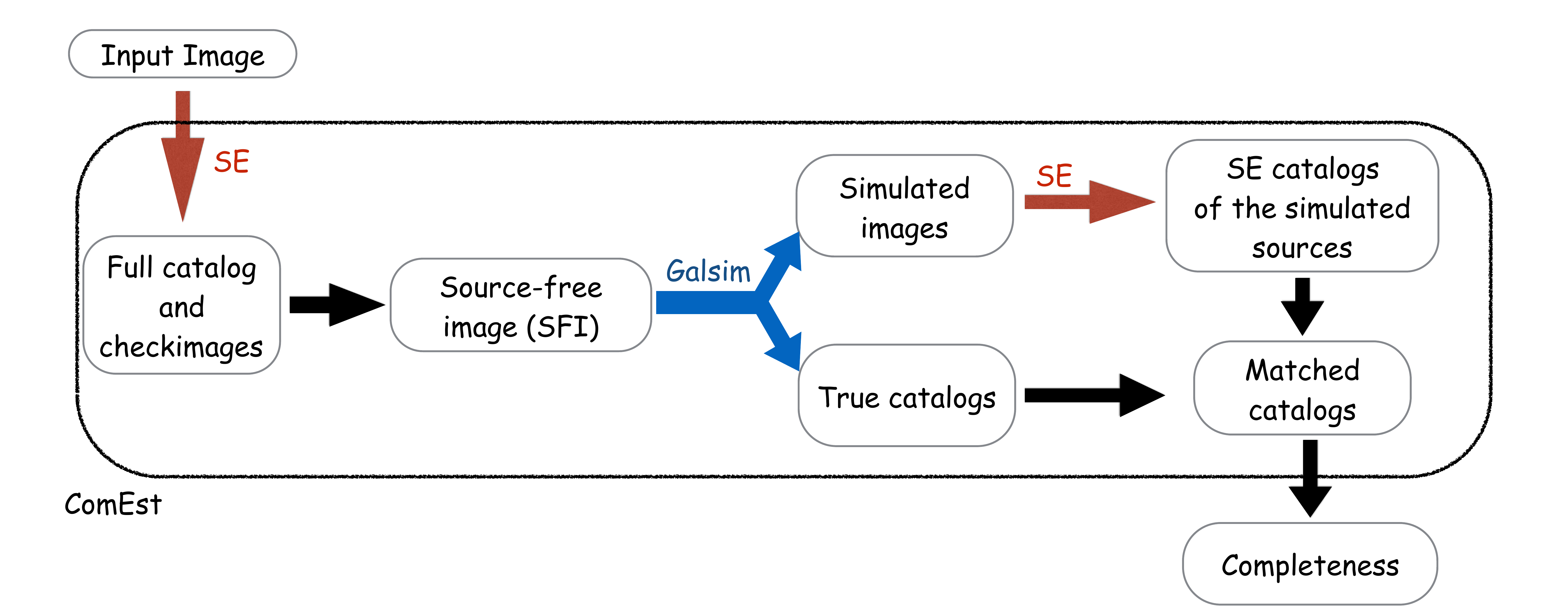}
}
\caption{
The workflow of \comest. The input image in the \fits\ format is the only input required by \comest. Once the input image is loaded, \comest\ runs \se\ to return the check-images and use the check-images to create the source-free image (SFI). Then \comest\ simulates various synthetic sources on SFI where \se\ is run on to detect the simulated sources. Finally, \comest\ matches the true catalog and the \se\ catalog to derive the completeness.
}
\label{fig:workflow}
\end{figure*}

The workflow of \comest\ is described as follows. 
For a given CCD image, \comest\ first runs \se\ to detect the observed sources and returns a set of standard outputs, especially the check-images.
Then, \comest\ removes the detected sources from the observed image and replaces them-- in the same positions-- by the background values estimated by \se.
In this way, \comest\ creates the source-free image (SFI) with the observed noise properties only.
Next, \comest\ puts various sources simulated by \galsim\ \citep{rowe15} on the SFI to create a set of simulated images.
After creating a set of simulated images, \comest\ re-runs \se\ again to detect the simulated sources.
Finally, \comest\ derives the completeness as a function of source flux and image position, by comparing the \se\ catalogs of the simulated sources and the true catalog used as the input to the simulation.
The workflow is shown in Figure~\ref{fig:workflow}. 
We describe the details of \comest\ in this section.

\subsection{Source detection and the source-free image (SFI)}
\label{sec:src_detection}

The Source Extractor \se\ \citep{bertin1996} is used as the source finder in \comest.
\se\ is a program which detects the sources on the images and performs the photometric measurements of them.
Since the first version in 1996, \se\ has been recognized as the standard tool for detecting sources and estimating the photometry on the images in  modern astronomy.
For a given image and the \se\ configuration, \se\ builds the catalog of the sources detected on the images and returns a set of diagnostic outputs, such as the check-images.
It is worth mentioning  that the performance of the \se\ depends on the  configuration file provided by the user. We refer readers to the \se\ manual\footnote{http://www.astromatic.net/software/sextractor} or \cite{dummies} for more details about the optimized usage of \se.

We stress that the observed image saved in the \fits\ format is the only input required by \comest.
The input image must contain the World Coordinate System (WCS) information in the header, which \comest\ extracts to get information about the astrometry.
\comest\ first runs \se\ on the input image to detect all the sources with  fluxes above the threshold assigned in the configuration file, then \se\ returns the output catalog and a set of check-images.
The goal of the first run of \se\ is to identify all the observed sources and estimate the background values on the image.
Once the sources are identified, \comest\ replaces the values in the pixels where the source is  located 
by randomly sampling the background values assuming the normal distribution with the mean and standard deviation in the check-images with the types of \texttt{BACKGROUND} and \texttt{BACKGROUND\_RMS}, respectively, estimated by \se.
%by the background values in the check-images with the types of \texttt{BACKGROUND} and \texttt{BACKGROUND\_RMS} estimated by \se.
In this way, \comest\ creates the source-free image (SFI) with only the observed noise properties present.
It is worth mentioning that the SFI image is-- by design-- free from the sources defined by \se\ configuration (e.g., $\mathtt{DETECT\_THRESH}$ or $\mathtt{PHOT\_AUTOPARAMS}$), therefore the sources undetected by \se\ (e.g., below the detection threshold) and the fluxes outside the extracted apertures remain the properties of the observed background on the SFI, resulting in negligible impact on the detections of other sources which are in the higher signal-to-noise regime (also see Section~\ref{sec:deriving_completeness}).
Running \se\ with the weight-map and other additional arguments is supported in \comest.

Alternatively, \comest\ can also remove only the sources which satisfy a certain criteria from the image and keep the others at the same positions.
Specifically, \comest\ can only remove the detected sources for which the fluxes (and sizes) are fainter (and smaller) than certain thresholds assigned by the \se\ arguments (\texttt{DETECT\_THRESHOLD} and \texttt{DETECT\_MINAREA}) and keep other observed sources on the image.
This is particularly useful when one wants to estimate the impact on completeness from the area masked by the relatively brighter and bigger sources on the image \citep{umetsu11}.

\subsection{Simulating sources}
\label{sec:simulation}

\comest\ generates the simulated images by placing the simulated sources on the SFI, which only keeps the artifacts and noise properties observed in the image.
The python module of the simulation toolkit-- \galsim-- is used as the central engine for simulating sources in \comest.
Two kinds of simulated sources are supported in \comest; the first one is the point sources (stellar-like sources) and the second one is the galaxies.
While simulating the sources (point source or galaxy), each source is randomly placed on the SFI with the magnitude (or the flux) uniformly distributed in the interval assigned by the user.
It is important that the well calibrated ZP of the input image must be provided to \comest\ for  simulating the sources with  unbiased photometry.

For simulating the point sources, \comest\ simply convolves each source with the Point Spread Function (PSF) using the assigned Full Width at Half Maximum (FWHM).
On the other hand, \comest\ simulates galaxies-- convolved with the PSF-- with the user-defined inputs for the half-light radius, major-to-minor axis ratio, and the position angle.
In addition, \comest\ currently provides two kinds of schemes for simulating galaxies: one can construct each galaxy by mixing the bulge and disk components with certain fractions, or one can also directly simulate the galaxies re-sampled from the galaxy catalog observed by the COSMOS survey \citep{ilbert09, rowe15}.

\comest\ simulates the sources on the SFI according to the pre-determined projected number density on the sky
\footnote{
In addition to simulating galaxies with the pre-determined projected number density, \comest\ also provides an utility (\texttt{comest.AnalysisSEcat}) to produce the simulated galaxy catalogs based on the magnitude distribution in the \se\ catalog extracted from the observed image.
In this way the source detection affected by the blending can be addressed.
}.
The number density of the simulated sources is assigned by users to avoid the blending problem in crowded fields.
In this way, \comest\ can avoid the incompleteness of the source detection, which is actually due to blending in the crowded field instead of the sky noise observed in the image. 
To achieve  better statistics in deriving the completeness, \comest\ can generate multiple simulated images with the same simulation scheme described above.
Simulating multiple images can be expedited by running \galsim\ using multiple threads, which is currently supported in \comest.
In the end, \comest\ returns the simulated images and the true catalogs which contain the input parameters for simulating the sources.

\subsection{Deriving completeness and purity}
\label{sec:deriving_completeness}

After simulating the sources on the SFI, \comest\ re-runs \se\ again on the simulated images to generate the \se\ output catalogs.
It is important that the same configuration, which is used in the first run of \se\ in \comest\ (see Section~\ref{sec:src_detection}), is also used for detecting the simulated sources, in order to have the derived estimates (e.g., the completeness of the detection) which are estimated from the identical process of the source extraction.

It is important to note that \comest\ assumes that the undetected sources-- which are present on the observed image but are not detected in the first run of \se-- remains undetected on the SFI with the same \se\ configuration and-- additionally-- they have negligible impact on the detectability and photometric measurement of the simulated sources higher than the detection threshold.
It is a reasonable assumption for the majority of cases where the source detections are not dominated by the background, i.e., the signal-to-noise of the source is well above the detection threshold.
On the other hand, the undetected sources near the confusion limit (e.g., the signal-to-noise ratio near to the $\mathtt{DETECT\_THRESH}$ in \se) contribute additional noise to the source detection on the SFI, hence the completeness estimates are likely to be biased low near the detection threshold, as a caveat of \comest.
Therefore, the users should pay attention to the \se\ configuration based on the magnitude range of interest where the completeness or purity properties are estimated.

After the source detection on the SFI, \comest\ matches the simulated sources detected by \se\ to the sources in the input catalog.
Specifically, for each detected source \comest\ first looks for the counterparts in the input catalog within the positional tolerance set by the users (default value is one FWHM), followed by the procedure of magnitude filtering such that the measured and input magnitudes are within the magnitude threshold (default is 1.5~mag).
That is, a matched pair has to be within the positional and magnitude tolerance.
If there are multiple objects within the positional tolerance in the first phase of matching, then the one within the magnitude threshold and with the closest magnitude difference is considered to be the matched object.
After visually inspecting the matched pairs, we find that majority of the matched pairs do not have multiple counterparts in the position matching for the number densities seen on the realistic sky, and adding the magnitude filtering after the positional matching can effectively remove the spurious matched pairs for the extreme case of $\approx200$~galaxies/arcmin$^2$. 

If the multiple simulated images are produced upstream, \comest\ will automatically merge the \se\ catalogs and the true catalogs of all the simulated images. 

Once the output catalogs of the simulated sources are built, \comest\ estimates the completeness as a function of magnitude (Equation~\ref{eq:derive_commag}) by comparing the fraction of the input sources recovered by \se.
\begin{eqnarray}
\label{eq:derive_commag}
\fcom(m) &= &\frac{N_{\mathrm{se}}(m)}{N_{\mathrm{true}}(m)} \, ,
\end{eqnarray}
where $\fcom(m)$, $N_{\mathrm{se}}(m)$ and $N_{\mathrm{true}}(m)$ are the completeness, number of simulated sources re-covered by \se\ and the number of  input sources in the simulation, respectively, at a  given magnitude  $m$.
In addition, \comest\ can estimate the completeness as a function of position on the input image at a given magnitude. i.e.,
\begin{eqnarray}
\label{eq:derive_comxy}
\fcom(x, y, m) &= &\frac{N_{\mathrm{se}}(x, y, m)}{N_{\mathrm{true}}(x, y, m)} \, ,
\end{eqnarray}
where $\fcom(x, y, m)$, $N_{\mathrm{se}}(x, y, m)$ and $N_{\mathrm{true}}(x, y, m)$ are the completeness, number of simulated sources re-covered by \se\ and the number of the input sources in the simulation, respectively, at the magnitude of $m$ and the coordinate $(x,y)$ of the input image.
Precisely, we derive the spatial distribution of number counts (e.g., $N_{\mathrm{se}}(x_{i}, y_{j}, m_{k})$ with the running indices $i$, $j$, $k$) by binning the boxes ($|x - x_{i}| < \Delta~x / 2$ and $|y - y_{j}| < \Delta~y / 2$) in the Cartesian coordinates 
%(i.e., constructing the two dimensional histogram)
based on the sources in the magnitude range of $|m - m_{k}| < \Delta~m/2$, $\Delta~x$, $\Delta~y$ and $\Delta~m$ are respectively the bin width of $x$-, $y$- coordinates and magnitude set up by the users.

On the other hand, \comest\ can estimate the purity of the source detection in a similar manner.
By comparing the catalogs produced by running \se\ on the simulated images and the true catalogs used as inputs to the simulation, the purity of the source detection is derived as the fraction of sources detected by \se\ which are also present in the input catalog of the simulation.
Specifically, the purity of the source detection as a function of magnitude $\fpur(m)$ and a function of position on the image give a magnitude cut $\fpur(x,y,m)$ is defined as follows. i.e., 
\begin{eqnarray}
\label{eq:derive_pur}
\fpur(m) &= &\frac{N_{\mathrm{se}}(m)}{N_{\mathrm{se}}(m) + N_{\mathrm{false}}(m)} \, \\
\fpur(x, y, m) &= &\frac{N_{\mathrm{se}}(x, y, m)}{N_{\mathrm{se}}(x, y, m) + N_{\mathrm{false}}(x, y, m)} \, ,
\end{eqnarray}
where $N_{\mathrm{false}}$ is the number of the false detections-- the sources detected by \se\ on the SFI which are not present in the true catalogs.

By defining the detection rate \fdr\ as the ratio of the numbers of all detected sources to the input sources, the above equations can be simply connected as Equation~\ref{eq:dr}.
That is, the detection rate contaminated by the false detection (Type I error or false positive) is quantified as the completeness of the source detection.
\begin{eqnarray}
\label{eq:dr}
\fdr    &\equiv    &\frac{ N_{\mathrm{se}} + N_{\mathrm{false}} }{ N_{\mathrm{true}} } \, , \nonumber \\ 
\fcom    &\equiv    &\fpur \times \fdr \, .
\end{eqnarray}

For calculating the completeness and purity, users can use a masked map to account for the masked area (e.g., by bright and saturated stars) which is present in the input image.
If the pixel is masked, then it would not be included in the calculation.
Note that the completeness and purity estimated by \comest\ depend on the configurations which are used for simulating the point sources and galaxies. It  is therefore  important to estimate the completeness by requiring realistic configurations (e.g., the half-light radius or FWHM) in the simulation.

Since \comest\ essentially re-detects the simulated sources and measures the photometry from end to end, in principle \comest\ can also be used to test the photometry performance (e.g., the ZP calibration) of the observed image by investigating the simulated sources detected by \comest. 
However, calibrating the photometry measurement is beyond the scope of this paper and we do not delve into it further.

%%%%%%%%%%%%%%%%%%%%%%%%%%%%
%
%  Example and Results
%
%%%%%%%%%%%%%%%%%%%%%%%%%%%%

\section{Example and Results}
\label{sec:example_results}

We demonstrate the usage of \comest\ in Section~\ref{sec:example} and present the results in  Section~\ref{sec:results}.

\subsection{Example}
\label{sec:example}

\begin{figure}
\centering
\resizebox{0.5\columnwidth}{!}{
\includegraphics[scale=1.0]{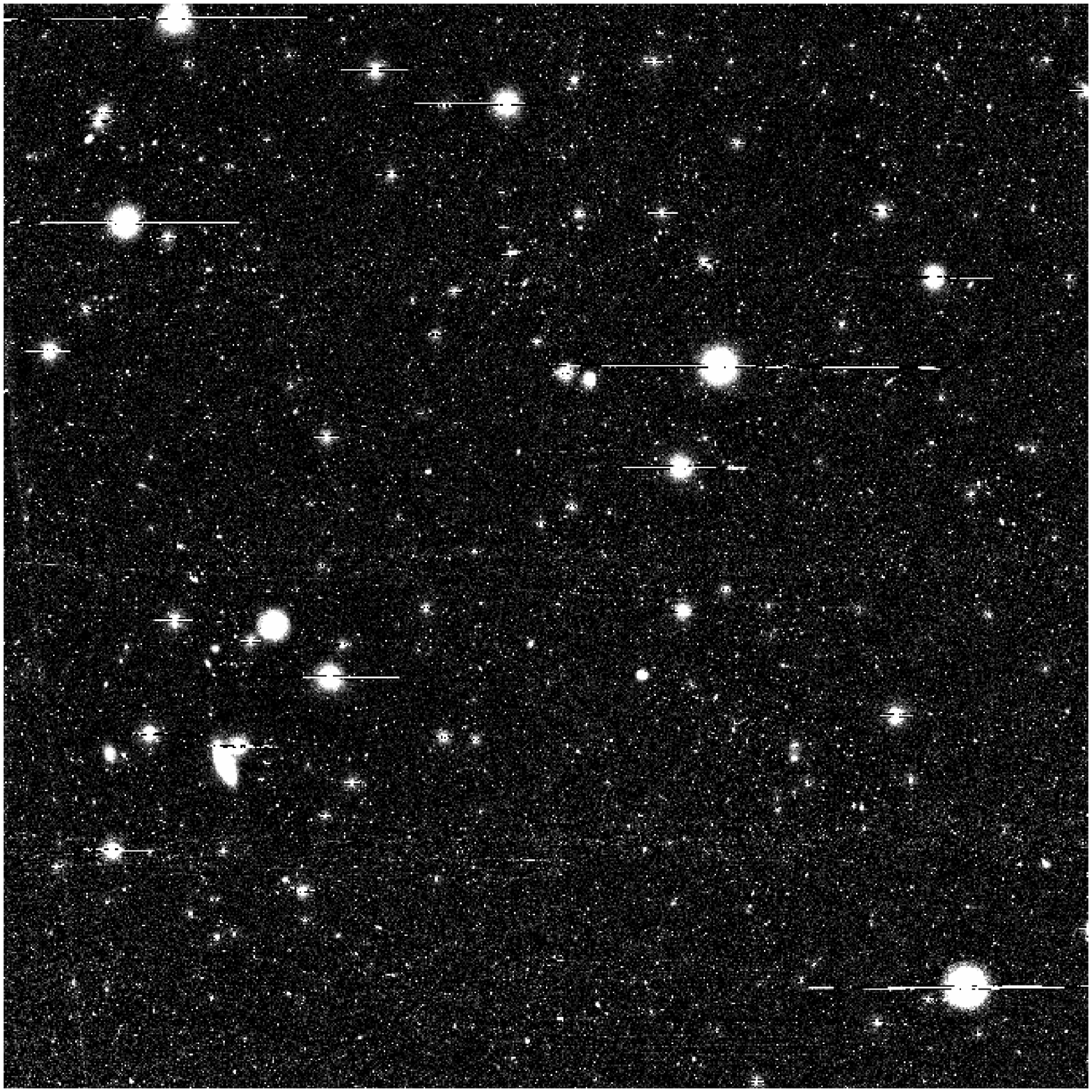}
}
\caption{
The image used as the example in this work.
This image is the $i$-band imaging of the footprint $2329+0012$ observed by the Blanco Cosmology Survey, and the field of view is $\approx 35.5 \times 35.5$~arcmin$^2$ (corresponding to $0.35$~deg$^2$) with the pixel scale of $0.26$ acrsec/pixel.
Note that the edge effects around the bright stars and the signatures of the chip gaps appear on Figure~\ref{fig:compurxy}, especially for the completeness maps.
}
\label{fig:original_img}
\end{figure}

The usage of \comest\ is described in detail in this section.
Specifically, we demonstrate the usage of \comest\ by running \comest\ on the footprint of BCS$2329+0012$ observed by the Blanco Cosmology Survey \citep{desai12} as an example.
We show the example image in Figure~\ref{fig:original_img}.
We assume that the observed image is called \texttt{example\_img.fits} in the current directory (\texttt{CWD}).
Moreover, all the output files produced by \comest\ will be saved in the directory called \texttt{outdir\_example}.
\begin{lstlisting}[language=Python, firstline=13, lastline=16]
# The absolute path to the input image
path2img        =       os.path.join( CWD, "example_img.fits" )
# The absolute path to the output directory
path2outdir     =       os.path.join( CWD, "outdir_example" )
\end{lstlisting}
It is also required by \comest\ to set the ZP of photometry \texttt{img\_zp} (in  units of magnitude), the pixel scale \texttt{img\_pixel\_scale} (in  units of arcsec/pixel) and the FWHM of the observed image \texttt{img\_fwhm} (in the unit of arcsec).
\begin{lstlisting}[language=Python, firstline=17, lastline=20]
# The image information
img_zp          =       31.7
img_pixel_scale =       0.26
img_fwhm        =       0.9
\end{lstlisting}

After the basic configuration is set (see above), we import the \comest\ package and then load  the input image with the required information.
\begin{lstlisting}[language=Python, firstline=25, lastline=33]
# Import comest
import comest
# Load in image
fits_image_example  =   comest.ComEst.fitsimage(
                        path2img = path2img,
                        path2outdir = path2outdir,
                        img_zp = img_zp,
                        img_pixel_scale = img_pixel_scale,
                        img_fwhm = img_fwhm)
\end{lstlisting}
Once the image is loaded, we first run \se\ on the input image to detect all the observed sources and then create the output catalog and check-images.
The output files of the first \se\ run are associated with the branch name of \texttt{full} (i.e., a set of output files with the file names containing \texttt{full}).
After the first run of \se, we can create the SFI (see Section~\ref{sec:src_detection}) with the file name of \texttt{full.identical.srcfree.fits}.
Optionally, \comest\ can create the ``BnB'' image which only contains the observed ``Bright and Big'' sources, with their SNR and size  larger than the assigned threshold (see Section~\ref{sec:src_detection}).
All the output files at this stage are saved in the directory of \texttt{outdir\_example}.
\begin{lstlisting}[language=Python, firstline=38, lastline=45]
# Run SExtractor for the first time
fits_image_example.RunSEforAll()
# Create the SFI and save it.
fits_image_example.SaveSrcFreeFits()
# Optional: run SExtractor to detect BnB sources
fits_image_example.RunSEforBnB()
# Optional: create the BnB image and save it.
fits_image_example.SaveBnBFits()
\end{lstlisting}

Next, we simulate the sources on the SFI image using the simulation toolkit \galsim\ (see Section~\ref{sec:simulation}).
For the purpose of illustration, we only demonstrate the case of simulating the galaxies which consist of bulge and disk components.
The same configuration scheme can be applied for simulating point sources and re-sampling the galaxies from the COSMOS catalog.
The simulated galaxies are uniformly sampled in the user-defined interval of magnitude, half-light radius, fraction of bulge component, minor-to-major axis ratio, and the position angle on the sky.
In this case we use the default values of \comest\ that the half-light radius and minor-to-major axis ratio span a range of $(0.35'', 0.75'')$ and $(0.4, 1.0)$, respectively.
\comest\ can simulate multiple images with the same configuration scheme in parallel by setting the arguments \texttt{nsimimages} and \texttt{ncpu}.
For example, one can simulate 10 images using 5 CPU cores by setting \texttt{nsimimages = 10} and \texttt{ncpu = 5}.
The number density of the simulated sources per arcmin square (\texttt{ngals\_arcmin2}) can be provided by the users.
It is worth mentioning that one should pay attention to the number density of the simulated sources because the completeness of the source detection will be inevitably affected by blending in the crowded field simulated by \comest.
In this example, we set the number density of the simulated sources to be 15 per arcmin square (\texttt{ngals\_arcmin2 = 15.0}).
In this simulation, all the output files (simulated images and true catalogs) are associated with the branch name of \texttt{buldisk} (by setting \texttt{sims\_nameroot = "buldisk"}).
\begin{lstlisting}[language=Python, firstline=50, lastline=57]
# Simulate galaxies (Bulge + Disk)
fits_image_example.BulDiskLocator(
                   path2image = fits_image_example.path2outdir + "/full.identical.srcfree.fits",
                   nsimimages = 10,
                   ngals_arcmin2 = 15.0,
                   ncpu       = 5,
                   sims_nameroot = "buldisk")
\end{lstlisting}
After the simulation tasks, we re-run \se\ on the 10 simulated images associated with the branch name of \texttt{buldisk} and output the resulting catalogs merged from the 10 \se\ runs. 
This is done by the following command.
\begin{lstlisting}[language=Python, firstline=61, lastline=62]
# Run SExtractor on the simulated images with the branch name of "buldisk"
fits_image_example.RunSEforSims(sims_nameroot = "buldisk")
\end{lstlisting}
Four catalogs associated to the branch name (i.e., \texttt{buldisk} for this case) are outputted, as listed in Table~\ref{tab:def_cat}.

In the end, using the output catalogs listed in Table~\ref{tab:def_cat} we derive the completeness of the source detection as a function of magnitude (Equation~\ref{eq:derive_commag}) and CCD position (Equation~\ref{eq:derive_comxy}) by invoking the following command.
\begin{lstlisting}[language=Python, firstline=68, lastline=69]
# Derive the completeness of the source detection for images of the branch name "buldisk"
fits_image_example.DeriveCom(sims_nameroot = "buldisk")
\end{lstlisting}
The completeness of the source detection is derived as the ratio of the source counts re-detected by \comest\ (\texttt{*.matched\_pairs.cat.fits}) to the source counts in the input catalog of the simulation (\texttt{*.merged\_true.cat.fits}) for a given magnitude cut and CCD position.
On the other hand, \comest\ can also derive the purity of the source detection-- which is defined by the ratio of the simulated source counts re-detected by \se\ (\texttt{*.sims.sex.matched\_pairs.cat.fits}) to the source counts of all detections (\texttt{*.sims.sex.matched\_pairs.cat.fits} + \texttt{*.sims.sex.ghost.cat.fits}).
This can be done by invoking
\begin{lstlisting}[language=Python, firstline=71, lastline=72]
# Derive the purity of the source detection for images of the branch name "buldisk"
fits_image_example.DerivePur(sims_nameroot = "buldisk")
\end{lstlisting}
We present the results in the following section.

\begin{table}[h]
\centering
\caption{
The output catalogs of running \se\ on the simulated images associated with the branch name \texttt{buldisk} by executing the method of \texttt{RunSEforSims}.
}
\label{tab:def_cat}
\resizebox{\columnwidth}{!}{
\begin{tabular}{ll}
\hline
Name & Definition \\
\hline\hline
\texttt{buldisk.sims.sex.merged\_true.cat.fits} & The merged true catalog of the simulated sources. \\
\texttt{buldisk.sims.sex.matched\_pairs.cat.fits} & The merged catalog of the simulated sources re-detected by \se. \\
\texttt{buldisk.sims.sex.unmatched.cat.fits} & The merged catalog of the simulated sources which are not re-detected by \se. \\
\texttt{buldisk.sims.sex.ghost.cat.fits} & The merged catalog of the sources which are detected by \se\ \\
 &but are not in the true catalog (i.e., the false detection) \\
\hline
\end{tabular}
}
\end{table}

\subsection{Results}
\label{sec:results}

The simulated images and the performance of photometry estimated by \se\ are presented in Section~\ref{sec:diag_outputs}.
We show the results in Section~\ref{sec:fcom_fpur}.

\begin{figure}[h]
\centering
\resizebox{0.7\columnwidth}{!}{ \includegraphics[scale=1]{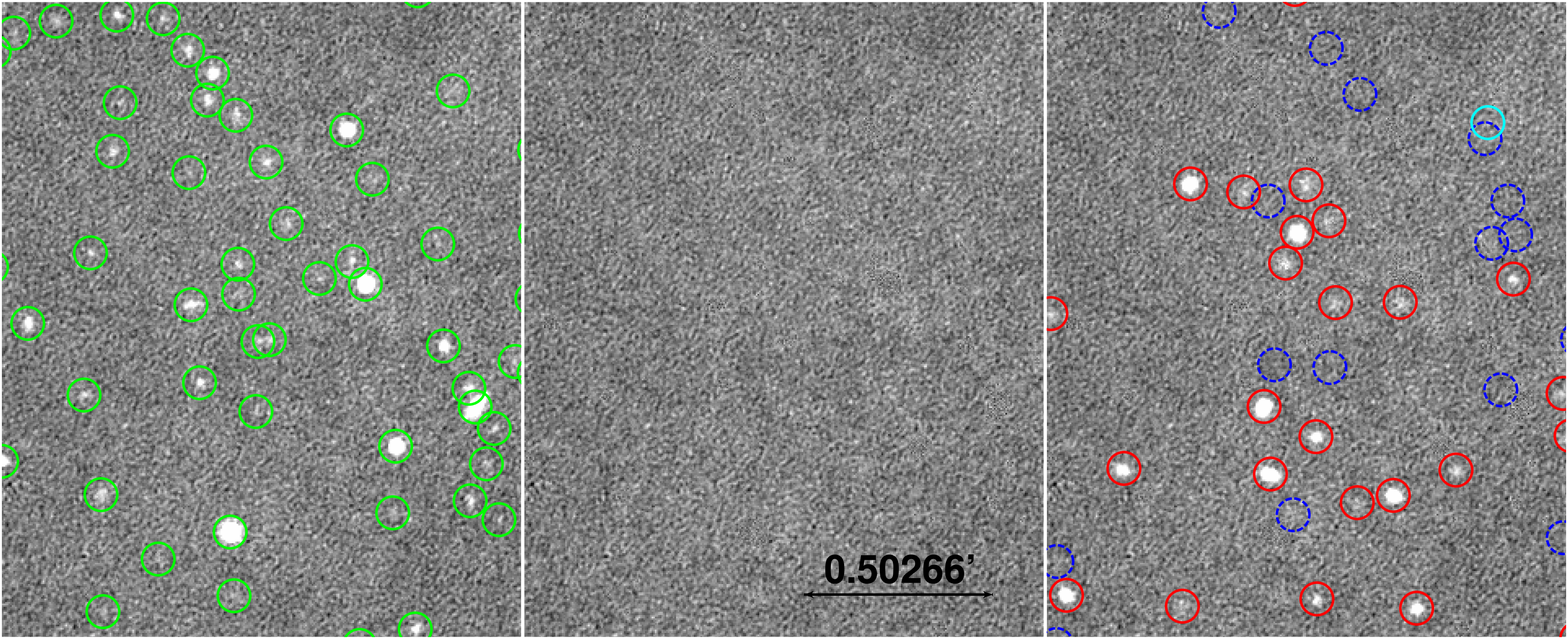} }\resizebox{0.3\columnwidth}{!}{ \includegraphics[scale=1]{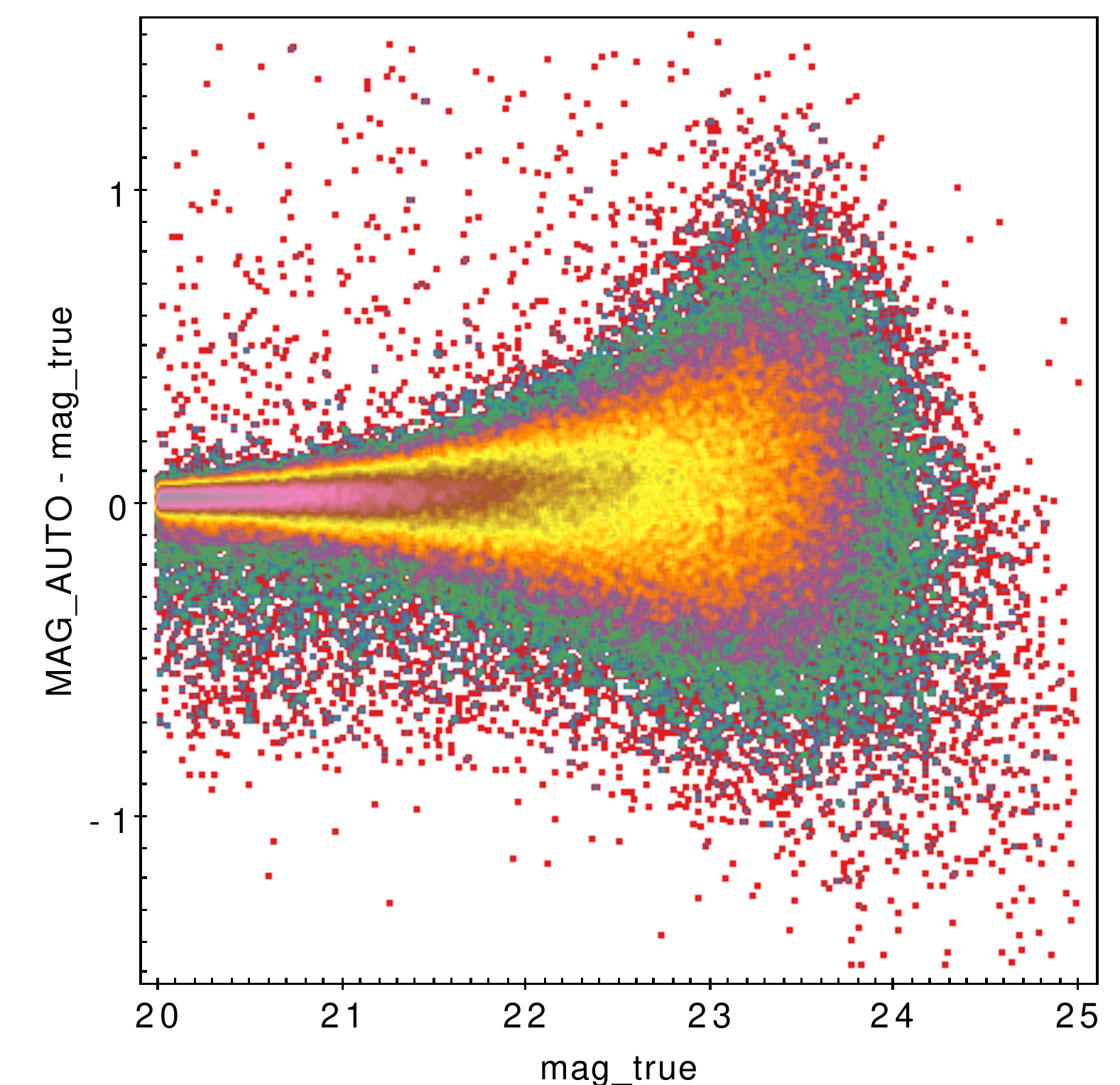}}
\caption{
The observed/simulated images and the derived photometry.
First panel: the observed image with the detected sources marked by green circles.
Second panel: the derived SFI (see Section~\ref{sec:src_detection}). No clear source is seen above the sky noise level.
Third panel: the simulated image by placing simulated galaxies (bulge + disk) on the derived SFI.
The red circles show the simulated sources which are re-detected by \se, while the blue-dashed circles indicate the non-detected sources simulated on the image.
The false detection (i.e., the sources are detected by \se\ on the simulated image but are not in the input catalog of the simulation) is marked by the cyan circle.
The scatter plot on the right is the \texttt{MAG\_AUTO} photometry estimated by \se\ compared to the input magnitude of the simulated sources. The $y$-axis is the \texttt{MAG\_AUTO} photometry estimated by \se, while the $x$-axis is the input magnitude \texttt{mag\_true} in the \galsim\ simulation. The density of the detected sources in the \texttt{MAG\_AUTO} v.s. \texttt{mag\_true} plane is plotted.
The black solid line shows the equality line.
}
\label{fig:magmag}
\end{figure}

\subsubsection{Diagnostic outputs}
\label{sec:diag_outputs}

\begin{figure*}[h]
\centering
\resizebox{\columnwidth}{!}{
\includegraphics[scale=1.0]{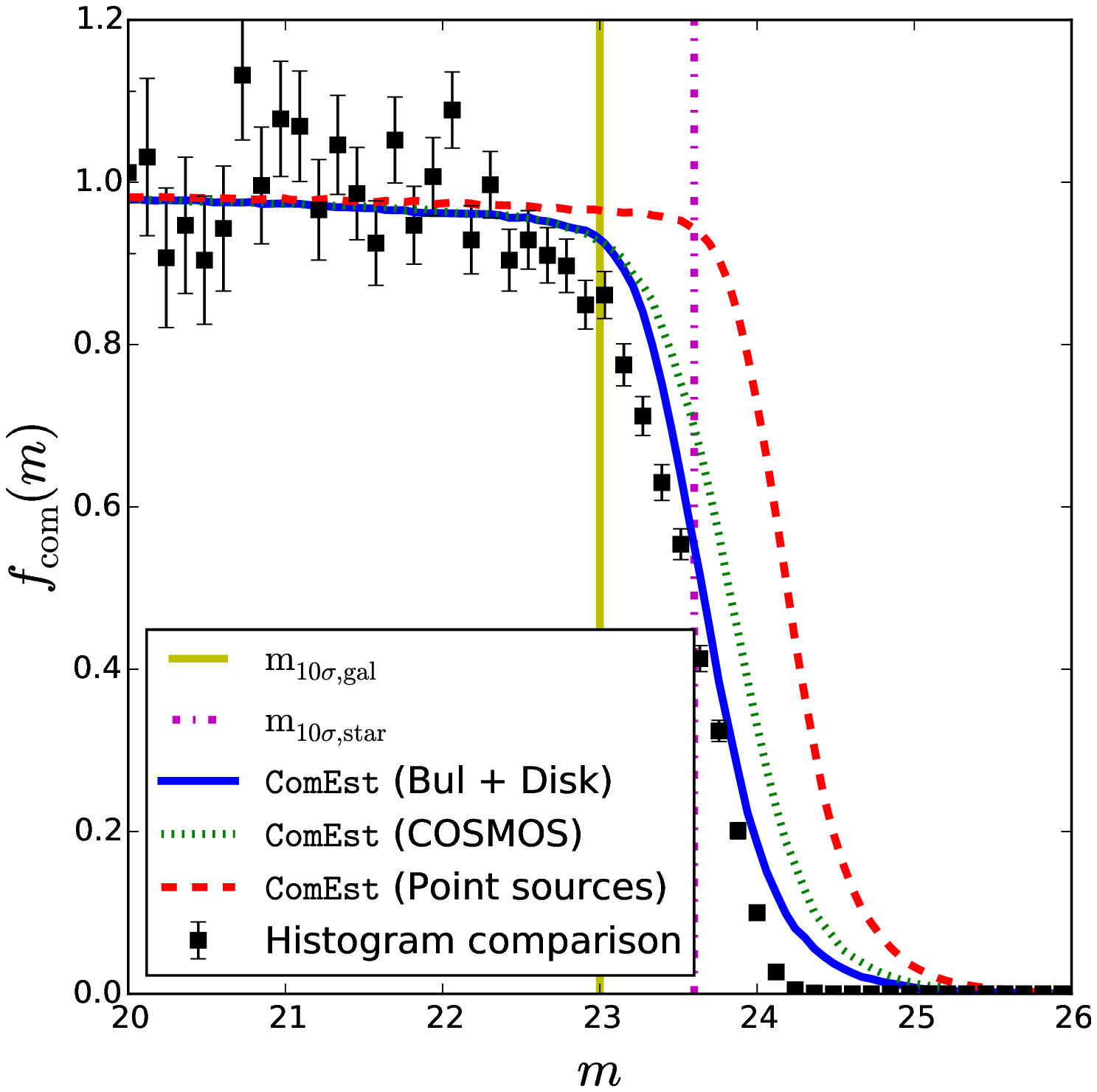}
\includegraphics[scale=1.0]{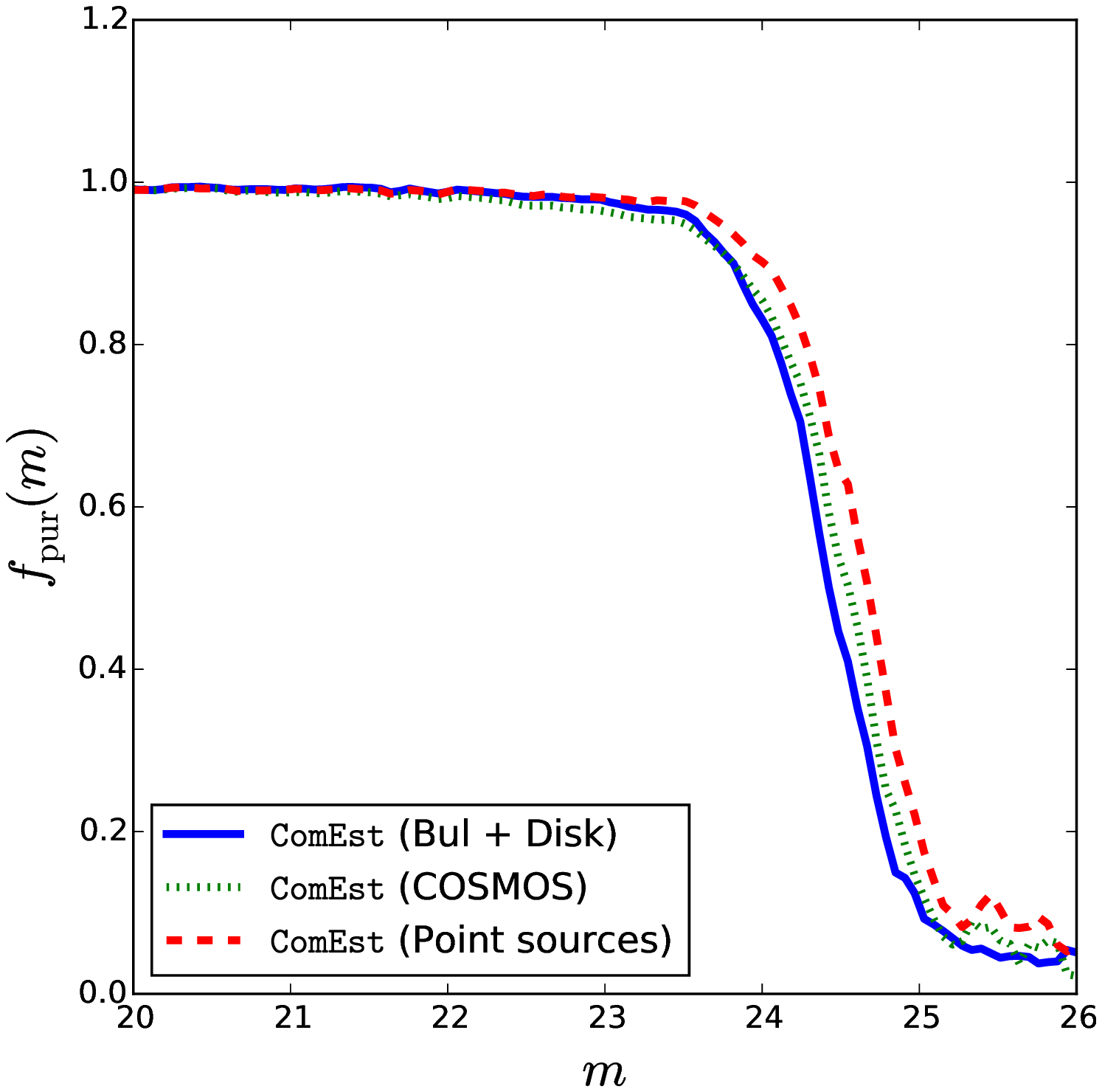}
}
\caption{The completeness and purity of the source detection as functions of magnitude.
\textit{Left panel}: the completeness estimated by \comest\ as a function of magnitude for the simulated galaxies consisting of bulge and disk components (blue solid line), galaxies re-sampled from the COSMOS catalog (green dotted line) and the simulated point sources (red dashed line).
The completeness derived from the histogram counts is shown by black squares.
The $10\sigma$ depth of galaxies and point sources are indicated by the yellow solid line and magenta dot-dashed line, respectively.
\textit{Right panel}: the purity estimated by \comest\ as functions of magnitude for the simulated galaxies consisting of bulge and disk components, galaxies re-sampled from the COSMOS catalog and the simulated point sources.
The color codes are the same as the left panel are used.
}
\label{fig:compur}
\end{figure*}
\begin{figure*}[h]
\centering
\resizebox{\columnwidth}{!}{
\includegraphics[scale=1.0]{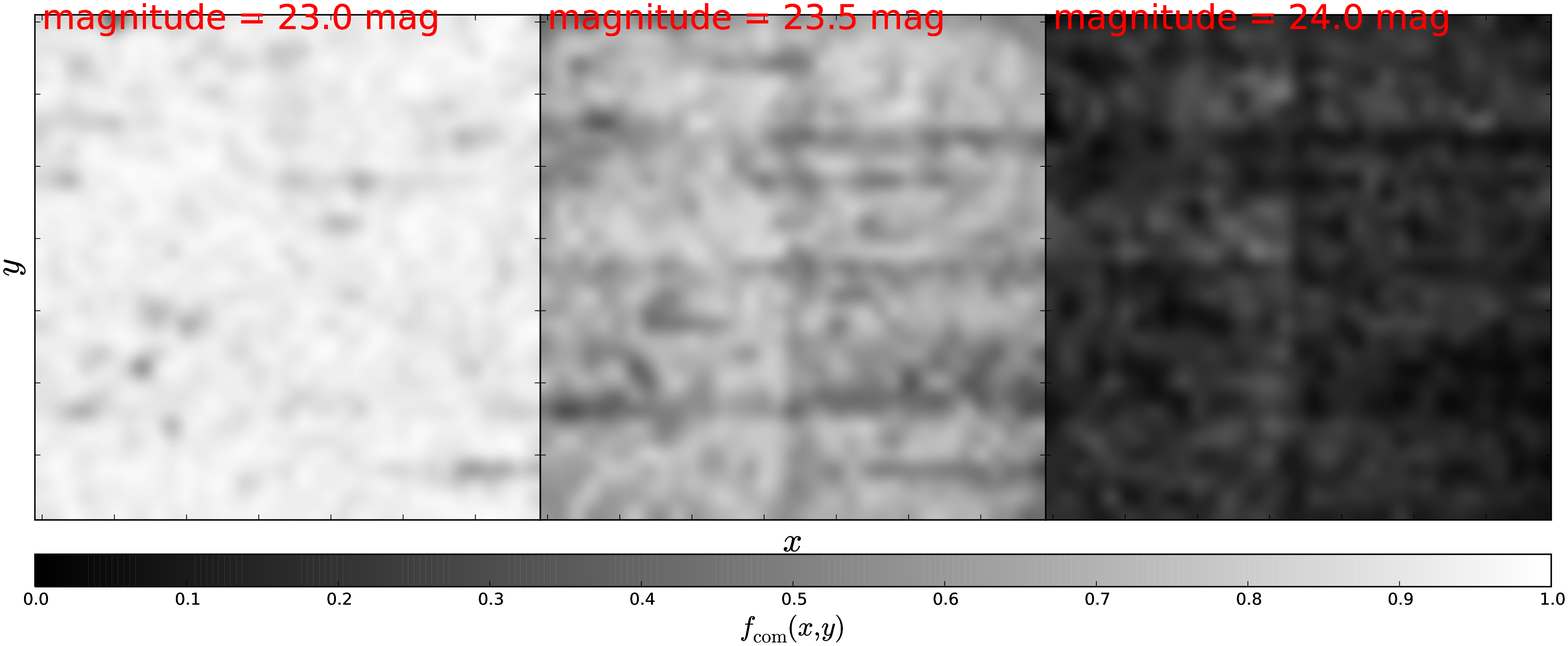}
}
\resizebox{\columnwidth}{!}{
\includegraphics[scale=1.0]{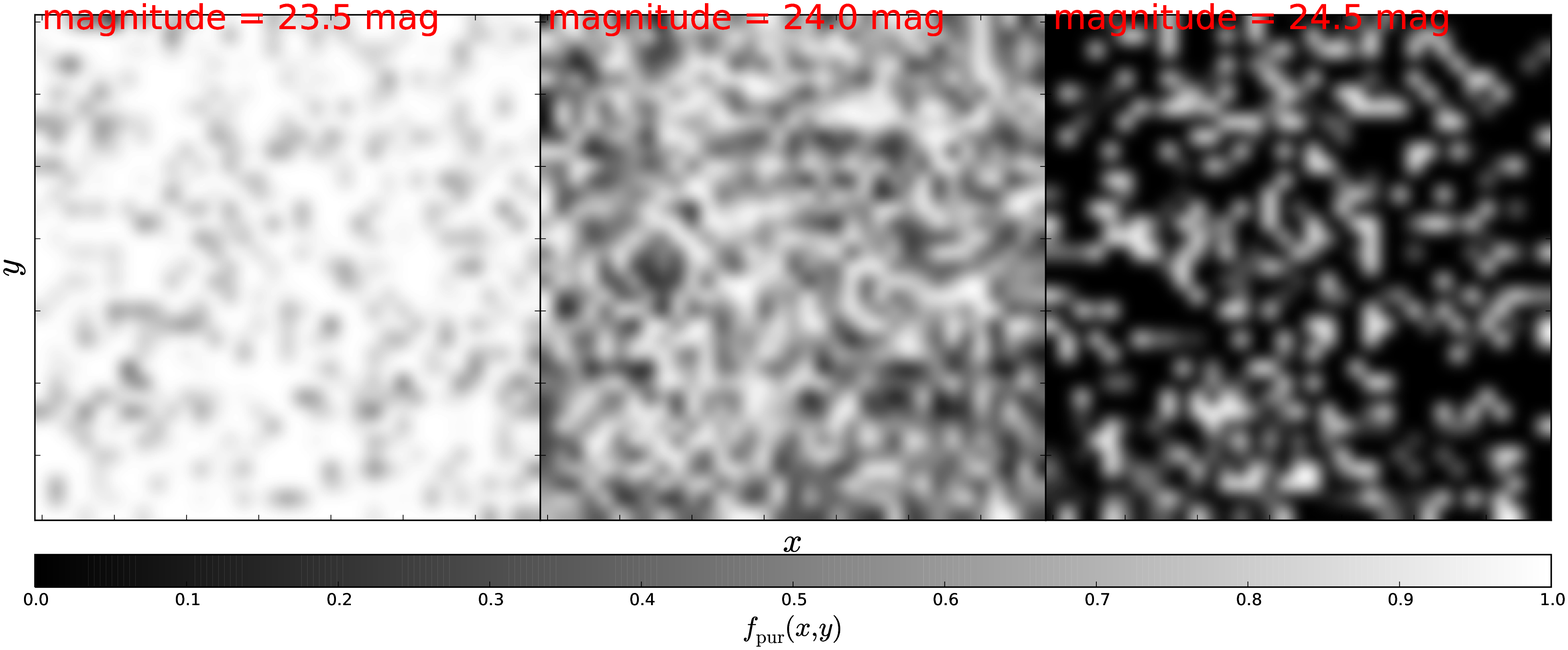}
}
\caption{
The completeness and purity maps estimated by \comest\ with the footprint size of $35.5'\times35.5'$.
\textit{Left panel}: the completeness map $(x, y)$ in the magnitude intervals centered on $m=23$~mag (left), $m=23.5$~mag (middle) and $m=24$~mag (right) with magnitude width $\Delta m=0.1$~mag.
The completeness of source detection is color-coded by the colorbar shown below the figures.
\textit{Right panel}: the purity map $(x, y)$ centered on $m=23.5$~mag (left), $m=24$~mag (middle) and $m=24.5$~mag (right).
The colorbar is the same as completeness map is applied.
}
\label{fig:compurxy}
\end{figure*}

The observed/simulated images and the derived photometry are shown in Figure~\ref{fig:magmag}.
In this simulation we simulate galaxies with magnitudes spanning a range between 20~mag and 27~mag.
The cutout of the observed tile BCS$2329+0012$ (the first panel), derived SFI (the second panel) and one out of 10 simulated images (the third panel) are shown in Figure~\ref{fig:magmag}.
As seen in Figure~\ref{fig:magmag}, the observed sources (green circles) are clearly detected by \se\ in the observed image and are then replaced by the estimated background noise to produce SFI, as shown in the second panel.
No clear signal above the observed sky noise is seen in the SFI image.
The galaxies consisting of the bulge and disk components are simulated on the derived SFI, as shown by the cutout image in the third panel.
The simulated sources detected by \se\ are marked by the red circles, while the blue-dashed circles indicate the sources, which are simulated but are not detected.
The false detection is marked by the cyan circle in the third panel.
In the fourth panel of Figure~\ref{fig:magmag} we show the comparison between the photometry estimated by \se\ and the input magnitude of the simulated sources.
Specifically, we compare the \texttt{MAG\_AUTO} photometry to the input magnitude \texttt{mag\_true}. 
The result suggests that the photometry performance is unbiased.
In addition, the clear drop of the source density in the plane of \texttt{MAG\_AUTO} v.s. \texttt{mag\_true} is seen, indicating that the source detection starts to suffer from the incompleteness at $\mathtt{mag\_true}\approx23.5~\mathrm{mag}$.

\subsubsection{Completeness and Purity}
\label{sec:fcom_fpur}

We first show the completeness $\fcom(m)$ and purity $\fpur(m)$ of the source detection as a function of magnitude in Figure~\ref{fig:compur}.
In the left panel of Figure~\ref{fig:compur}, we also show the completeness of point sources and galaxies re-sampled from the observed COSMOS catalog (see Section~\ref{sec:simulation}) estimated by \comest.
In addition, the limiting magnitudes calculated using the  $10\sigma$ depths for point source  and galaxies are taken from \cite{desai12} and the completeness of the source detection derived from the histogram counts (see Section~\ref{sec:introduction}) are also plotted for comparison.
The completeness derived by \comest\ agrees reasonably well for the case of galaxies, with the tendency that the magnitudes associated with the completeness levels derived by re-sampling the COSMOS catalog are slightly deeper (by $\approx0.1-0.15$~mag) than the results using the galaxies consisting of bulge and disk components and using the histogram count technique.
The $10\sigma$ depth of galaxies indicates the completeness at the level of $\approx92\percent$ and $\approx85\percent$ based on the estimates of \comest\ and histogram count technique, respectively.
On the other hand, the magnitudes which correspond to  the same completeness level for the point sources are overall fainter than the galaxies by $\approx1$~mag, which is consistent with the difference of the $10\sigma$ depth seen between the galaxies and point sources.
The purity of the source detection for point sources, galaxies consisting of the bulge and disk components and the galaxies re-sampled from COSMOS catalog are shown in the right panel of Figure~\ref{fig:compur}.
There is no significant difference of the purity seen among the simulated samples, indicating that the $50\percent$ completeness corresponds to the purity of $\approx94\percent$ and $\approx72\percent$ for galaxies and point sources, respectively.

We show the results of completeness and purity of the source detection estimated by \comest\ as functions of CCD position given the magnitude threshold in Figure~\ref{fig:compurxy}.
The completeness and purity maps are constructed by binning $1$ arcmin in both $x$ and $y$ direction and are showed in Figure~\ref{fig:compurxy} after the bicubic interpolation.
To get better statistics on a single cell basis in the magnitude and position space $(x,y,m)$, we simulate 100 images based on the derived SFI (see Section~\ref{sec:simulation}) in deriving the $\fcom(x,y,m)$ and $\fpur(x,y,m)$.
Note that the finner binning needs more simulated images to achieve enough statistics (i.e., the galaxy counts) in a single binned cell, hence the users should compromise between the desired resolution in the spatial variance and the running time spent on the image simulations.
For simplicity, we only show the results of galaxies consisting of bulge and disk components (the blue solid lines in Figure~\ref{fig:compur}), since the same qualitative picture is suggested by other cases estimated by \comest.
Specifically, we show the results of completeness (purity) as a function of CCD position given the magnitude cut in the intervals centering on $m = 23$~mag, $m = 23.5$~mag and $m = 24$~mag ($m = 23.5$~mag, $m = 24$~mag and $m = 24.5$~mag) with the width of $0.1$~mag in the upper (lower) panel of Figure~\ref{fig:compurxy}.
It is worth mentioning in Figure~\ref{fig:compurxy} that the edge effect around the bright stars seen in Figure~\ref{fig:original_img} is captured by \comest, while the chip gaps-- which are not easy to be visually inspected in Figure~\ref{fig:original_img}-- are clearly characterized in the completeness map.
These signatures are less prominent in the purity map at the very faint end (e.g., $\gtrsim 24.5$~mag) because of the fact that the detection rate (i.e., Equation~\ref{eq:dr}) is too low ($N_{\mathrm{se}} + N_{\mathrm{false}} \approx 0$) to have the meaningful estimates of the purity, causing the purity maps consisting of the cells with $1\times1$~arcmin$^2$ suffer from large poisson fluctuation.
The resulting completeness and purity maps of the sourced detection show reasonable uniformity in the context of the spatial properties of the tile BCS$2329+0012$.

%%%%%%%%%%%%%%%%%%%%%%%%%%%%
%
%  Conclusion
%
%%%%%%%%%%%%%%%%%%%%%%%%%%%%

\section{Conclusions}
\label{sec:conclusion}

We have presented a python package \comest\ to estimate the completeness and purity of the source extractor \se\ used for the optical/NIR imaging of point sources or galaxies, by utilizing the state-of-the-art simulation toolkit \galsim.
The observed image saved in the FITS format with a well-calibrated ZP is the only input required  by \comest.
\comest\ carries out the simulation and source detection directly on the observed image, which ensures that the information about the observational artifacts and noise properties are preserved while deriving the completeness and purity estimates.
\comest\ can simulate galaxies with various properties, such as half-light radius or minor-to-major axis ratio.
The running time of the simulations conducted by \comest\ is generally fast, it takes about $\approx 60$~seconds for simulating 15 galaxies/arcmin$^2$ on a 0.35 deg$^{2}$ footprint with the pixel scale of $0.26$ arcsec/pixel.
In addition, running multiple simulations in parallel is supported in \comest.
It is worth noting that \comest\ estimates the completeness and purity of the source detection as a function of magnitude and CCD position in a fully automated fashion with an easy-to-use syntax.
We expect that \comest\ would be extremely helpful in quantifying the systematics related to the completeness and purity for the analysis such as the galaxy clustering or the lensing magnification.

\comest\ is publicly available and can be downloaded from  https://github.com/inonchiu/ComEst, along with  full documentation and more detailed examples.

\section*{Acknowledgements}
\label{sec:acknowledgements}

This paper is dedicated to Chien-Ho Lin in Taiwan.
We acknowledge the support by the DFG Cluster of Excellence ``Origin and Structure of the Universe'', the DLR award 50 OR 1205 that supported I.~Chiu during his PhD project, and the Transregio program TR33 ``The Dark Universe''. 
The computations for \comest\ have been carried out on the computing facilities of the Computational Center for Particle and Astrophysics (C2PAP) and of the Leibniz Supercomputer Center (LRZ).

\clearpage
%%%%%%%%%%%%%%%%%%%%%%%%%%%%
%
%  Bibtex
%
%%%%%%%%%%%%%%%%%%%%%%%%%%%%

\section*{References}
\bibliographystyle{IEEEtranN}
\bibliography{ComEst}

\end{document}